\documentclass[preprint,onecolumn,nofootinbib]{revtex4}
\usepackage[colorlinks=true,linkcolor=blue,urlcolor=blue,filecolor=black,citecolor=red,pdfstartview=FitV,pdftitle={},pdfsubject={},pdfkeywords={},pdfpagemode=None,bookmarksopen=true]{hyperref}
\usepackage{graphicx}
\usepackage{amsmath}
\usepackage{amsfonts}
\usepackage{amssymb,ulem}
\usepackage{color,xcolor}%
\usepackage{CJK}
\usepackage{subfigure}
\usepackage{amsthm,amsmath,amssymb}
\usepackage{mathrsfs}
\usepackage{multirow}

\usepackage{dcolumn}
\usepackage{float}
\begin{document}
\title{Echoes from the Minkowski-core spacetime}
\author{Dan Zhang $^{1}$}
\thanks{danzhanglnk@163.com}
\author{Qin Tan$^{2}$}
\thanks{tanqin@hunnu.edu.cn}
\author{Guoyang Fu$^{1}$}
\thanks{FuguoyangEDU@163.com}
\author{Huajie Gong$^{2}$}
\thanks{huajiegong@hunnu.edu.cn}
\author{Jian-Pin Wu$^{1}$}
\thanks{jianpinwu@yzu.edu.cn}
\author{Qiyuan Pan$^{2}$}
\thanks{panqiyuan@hunnu.edu.cn}
\affiliation{$^1$\mbox{Center for Gravitation and Cosmology, College of Physical Science and Technology,} \mbox{Yangzhou University, Yangzhou 225009, China}\\
$^2$Department of Physics, Key Laboratory of Low Dimensional Quantum Structures and Quantum Control of Ministry of Education, \mbox{ Synergetic Innovation Center for Quantum Effects and Applications,} \mbox{ Hunan Normal University, Changsha, 410081, Hunan, China}}
	
\begin{abstract}


In this study, we construct a class of horizonless exotic compact objects (ECOs) with Minkowski-core, classifying them as either photon-sphere ECOs (PS-ECOs) or photon-sphere lacking ECOs (PL-ECOs) based on photon sphere topology. Time-domain analysis reveals that the dynamical evolution can be divided into three phases: the initial ringdown, the echo phase, and the final ringdown. The echo signals exhibit the periodic damping, with quantum effects significantly accelerating the echo dissipation and prompting an earlier transition to the long-lived mode-dominated phase. Furthermore, the QNM spectrum of the PS-ECO exhibits fundamentally different behavior from that of BHs—including the presence of long-lived modes and the absence of overtone outbursts—providing a clear spectroscopic signature distinguishing PS-ECOs from BHs.  This work is significant in providing new theoretical foundations and waveform features for identifying such quantum-corrected ECOs, contributing critically to the understanding of quantum gravity effects.

\end{abstract}

\maketitle


\section{Introduction}\label{sec-intro}

The landmark detections of gravitational waves (GWs) by the LIGO and Virgo interferometers have powerfully validated theoretical predictions of general relativity (GR) concerning black holes (BHs) \cite{LIGOScientific:2016vpg,LIGOScientific:2016aoc,LIGOScientific:2021usb,LIGOScientific:2020ibl,KAGRA:2021vkt}. The ongoing advancement of GW detection technology is opening new avenues for probing quantum gravity effects and detecting related physical phenomena such as GW echoes \cite{Lai:2025skp}.

To address the fundamental issue of spacetime singularities in GR, a class of regular black hole (RBH) models has been developed \cite{Bardeen:1968,Hayward:2005gi,Frolov:2016pav,Xiang:2013sza,Culetu:2013fsa,Culetu:2014lca,Rodrigues:2015ayd,Simpson:2019mud,Ghosh:2014pba,Ghosh:2018bxg,Li:2016yfd,Martinis:2010zk,Ling:2021olm,Bambi:2023try,Vagnozzi:2022moj,Lan:2023cvz}. Their phenomenological construction typically involves exotic matter that violates classical energy conditions, such as the inclusion of nonlinear electrodynamics sources. This has led to their consideration as potential manifestations of quantum gravity effects. Based on the asymptotic characteristics of their central regions, RBHs can be categorized into two distinct classes. The first class possesses a de Sitter core, where the central energy density approaches a constant value, as exemplified by Bardeen, Hayward, and Frolov BHs \cite{Bardeen:1968,Hayward:2005gi,Frolov:2016pav}. The second class features a Minkowski core,  where the central energy density vanishes asymptotically \cite{Xiang:2013sza, Culetu:2013fsa, Culetu:2014lca, Rodrigues:2015ayd, Simpson:2019mud, Ghosh:2014pba, Ghosh:2018bxg, Li:2016yfd, Martinis:2010zk, Ling:2021olm}. The Minkowski-core models exhibit a simpler curvature structure than their de Sitter-core counterparts, greatly simplifying the description of deep-core physics \cite{Simpson:2019mud}.

Numerous studies building upon the Minkowski-core BH model proposed in \cite{Ling:2021olm} have been conducted, as seen in \cite{Zhang:2024nny,Ling:2022vrv,Zeng:2023fqy,Zeng:2022yrm,Guo:2025zca}. These studies are dedicated to analyzing the model's distinctive features to uncover potential quantum gravity effects. A finding from quasi-normal mode (QNM) analyses is the pronounced outburst of overtones in this RBH compared to the Schwarzschild case \cite{Zhang:2024nny}. This feature is attributed to modifications in the near-horizon geometry induced by quantum gravity effects. Although this model shares the same large-scale behavior with the Hayward BH and some loop quantum gravity (LQG)-corrected BH, a similar phenomenon of overtone outburst is observed in its QNM spectrum \cite{Zhang:2024nny}. Furthermore, shadow imaging studies suggest that quantum correction parameters influence both the size and shape of the BH shadow. Specifically, shadows of Minkowski-core BH exhibit stronger deformation than those with a de Sitter core \cite{Ling:2022vrv}. Collectively, these findings offer promising avenues for distinguishing between different regular BH cores through future astronomical observations and for testing quantum gravity effects.

Recently, a study has revealed that for every category of RBHs, there exists a corresponding category of horizonless stars \cite{Carballo-Rubio:2022nuj,Borissova:2025msp}. Given mild conditions of non-negative gravitational energy (Misner–Sharp quasi-local mass) and a linear relation between this mass and the Arnowitt–Deser–Misner (ADM) mass, a RBH formed through the gravitational collapse may evolve into a horizonless ultracompact star. This implies that RBHs and horizonless compact stars are not entirely independent, but rather two manifestations governed by different parameters within the same theoretical framework. RBHs and horizonless ultracompact stars exhibit fundamental structural differences. Specifically, RBHs possess both inner and outer horizons but lack a singularity. In contrast, horizonless ultracompact stars are even more extreme, having no horizons entirely, and their surfaces locate deep inside the photon sphere \cite{Chirenti:2007mk,Cardoso:2019rvt,Carballo-Rubio:2022aed}. 

Given the persistent challenges in detecting BH event horizons and the close connection between the BH information paradox and the existence of event horizons, exploring extremely compact objects as alternatives to BHs has become an important topic in current theoretical research. The aforementioned horizonless compact stars can be regarded as a type of exotic compact object (ECO). Typical ECO models include wormholes \cite{Morris:1988tu,Damour:2007ap}, fuzzballs \cite{Mathur:2005zp}, gravastars \cite{Mazur:2001fv}, boson stars \cite{Schunck:2003kk}, and other configurations \cite{Barcelo:2010vc,Holdom:2016nek,Konoplya:2019nzp}. However, as shown in \cite{Cardoso:2016rao}, the detection of GW signals resembling those from the ringdown phase of a classical BH is insufficient to confirm the existence of an event horizon. Due to their ability to emulate key features of BHs, horizonless compact objects can produce GW signals during the early ringdown phase that are virtually indistinguishable from those of true BHs. GW echoes may serve as a distinctive signature of horizonless compact objects, providing new observational evidence for their identification \cite{Cardoso:2016oxy,Cardoso:2017cqb,Mark:2017dnq,Konoplya:2018yrp,LongoMicchi:2019wsh,Churilova:2019cyt,Bronnikov:2019sbx,DuttaRoy:2019hij,Chowdhury:2020rfj,Chowdhury:2022zqg}.

In this study, we construct a horizonless compact object models corresponding to the Minkowski-core RBH proposed in \cite{Ling:2021olm}. By studying electromagnetic perturbations in this spacetime, we analyze how quantum correction parameters influence the resulting GW echo phenomena. The structure of this paper is organized as follows. Section \ref{sec1} constructs the spacetime geometry and analyzes its features. Section \ref{sec2} addresses electromagnetic perturbations in this background. The resulting echo waveforms are studied in Section \ref{sec3}, followed by conclusions and discussion in Section \ref{sec4}. The Appendix \ref{method} outlines the numerical methods employed.

\section{Geometry of Minkowski-core model}\label{sec1}

In this section, we demonstrate that the RBH model proposed in Ref. \cite{Ling:2021olm} simultaneously describes both RBHs and horizonless ultracompact objects. We therefore designate this solution as the Minkowski-core spacetime, with the metric:
\begin{eqnarray} \label{metric}
ds^2=-f(r)dt^2+\frac{1}{f(r)}dr^2+r^2d\Omega^2\,, \ \ f(r)=1-\frac{2M}{r}e^{\frac{-\alpha_{0}M^x}{r^c}}\,,
\end{eqnarray}
where $\alpha_0$ quantifies quantum corrections and measures the deviation from the Schwarzschild spacetime. For $\alpha_0 = 0$, the metric reduces to the Schwarzschild case. The dimensionless exponents $x$ and $c$ govern the asymptotic behavior of the Kretschmann scalar curvature at the core. Notably, this RBH features an asymptotically Minkowski core ($r \to 0$), distinguishing it from de Sitter-core models (e.g., Hayward/Bardeen BHs). In this paper, we focus on the parameter choice $x=1$, $c=3$.

\begin{figure}[ht]
	\centering{
		\includegraphics[width=7.8cm]{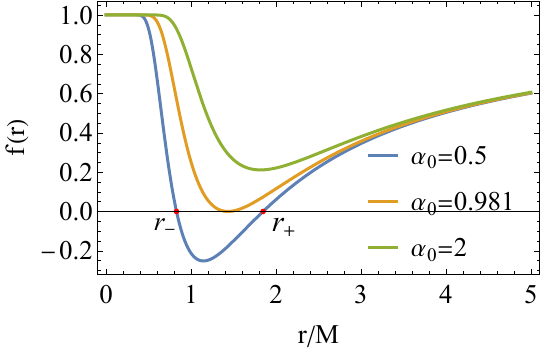}\hspace{0.05mm}
		\includegraphics[width=7.8cm]{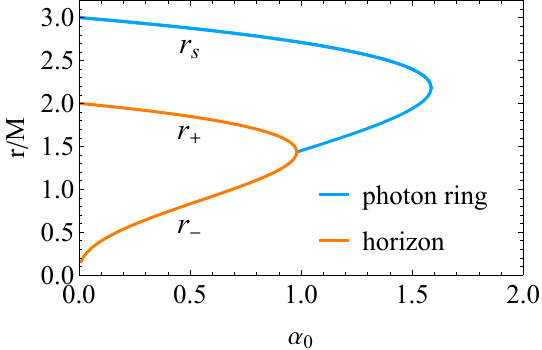}
		\caption{The left panel shows the behavior of the function $f(r)$ as a function of $r$. The right panel illustrates the variation of the horizon radius (yellow curve) and the photon sphere radius (blue curve) with respect to the quantum correction parameter $\alpha_{0}$.}
		\label{fig1}
	}
\end{figure}

The quantum correction parameter $\alpha_0$ characterizes the geometric properties of spacetime. As shown in the left panel of Fig.~\ref{fig1}, a small $\alpha_0$ yields a RBH with two horizons. At the critical value $\alpha_0^* \approx 0.981$, these horizons merge, forming an extremal BH. When $\alpha_0$ is larger than this critical value, the horizon disappears, revealing a horizonless compact object \cite{Carballo-Rubio:2022nuj}. Such objects are classified by the photon sphere topology as either photon-sphere ECOs (PS-ECOs) or photon-sphere lacking ECOs (PL-ECOs) \cite{Virbhadra:2002ju}. Fig.~\ref{fig1} depicts the evolution of the horizon structure and photon rings, while Table~\ref{tab_1} categorizes the $\alpha_0$ parameter space accordingly.

\begin{table}[H]
	\centering
    \begin{tabular}{lc}
    \hline \textbf{Type} & \textbf{$\alpha_0$ Range} \\
    \hline  \text{BH with two horizons} & 0$\leq \alpha_0 <  0.981$  \\
    \hline  \text{Extremal BH (single horizon)} & $\alpha_0^*\approx0.981$   \\
    \hline \text{PS-ECO} & $0.981 < \alpha_0 < 1.586$  \\
    \hline \text{PL-ECO} & $\alpha_0 \geq 1.586$  \\
    \hline
    \end{tabular}
    \\
\caption{ Classification of the Minkowski core spacetime. The evolution from RBHs to ECOs is governed by the quantum correction parameter $\alpha_0$, with critical transitions at $\alpha_0^* \approx 0.981$ and $\alpha_0 \approx 1.586$.}  \label{tab_1}
\end{table}


\section{Electromagnetic Field Perturbations}\label{sec2}

In this work, we model the evolution of a test electromagnetic field as a proxy for gravitational perturbations. This approach captures key qualitative features of GW echoes. The radial equations of motion (EOMs) for electromagnetic perturbations are governed by:
\begin{eqnarray}\label{eq_ele}
	\frac{1	}{\sqrt{-g}}\partial_\mu\left(\sqrt{-g} F_{\gamma\sigma}g^{\gamma\mu}g^{\sigma\nu}\right)=0\,,
\end{eqnarray}
where $F_{\mu\nu}=\partial_\mu A_\nu-\partial_\nu A_\mu$ represents the 
strength of the electromagnetic field and $A^{\mu}$ denotes the vector potential. 

Exploiting spherical symmetry, we decompose the electromagnetic field via the vector spherical harmonics \cite{Rosa:2011my,Cannizzaro:2020uap}:
\begin{eqnarray}\label{sph_harmonics}
Z_\mu^{(1)lm}&=&\left[1,0,0,0\right]Y^{lm}\,,  \\
Z_\mu^{(2)lm}&=&\left[0,f^{-1},0,0\right]Y^{lm}\,, \\
Z_\mu^{(3)lm}&=&\frac{r}{\sqrt{l(l+1)}} \left[0,0,\partial_\theta,\partial_\phi\right]Y^{lm}\,,  \\
Z_\mu^{(4)lm}&=&\frac{r}{\sqrt{l(l+1)}}\left[0,0,\frac{\partial_\phi}{\sin\theta},-\sin\theta\partial_\phi\right]Y^{lm}\,,
\end{eqnarray}
with $Y^{\ell m}$ being the scalar spherical harmonics, $l$ and $m$ represent the harmonics indices. The vector field separates into parity-distinct components under transformation $(\theta,\phi) \to (\pi-\theta,\phi+\pi)$:
\begin{eqnarray}\label{A_mu}
A^{\mu}=A_{odd}^{\mu}+A_{even}^{\mu}.
\end{eqnarray}

Following the procedure in \cite{Ruffini,Cardoso:2003pj,Fu:2023drp}, we recast Eq.~\eqref{eq_ele} into the Schr\"{o}dinger-like form:
\begin{equation}\label{eq_Sch}
	\frac{\partial ^2 \psi^{em}}{\partial t^2} -\frac{\partial ^2 \psi^{em}}{\partial r_*^2} + V_{eff}(r) \psi^{em}=0\,,
\end{equation}
where $\psi^{em}=a^{lm}$ for the odd parity, and $\psi^{em}=\frac{r^2}{l(l+1)}(\partial_th^{lm}-\partial_r d^{lm})$ for the even parity. The effective potential is given by: 
\begin{equation}\label{V_eff}
	V_{eff}(r)=f(r) \frac{l(l+1)}{r^2}.
\end{equation}

For BH spacetimes, the effective potential $V_{\mathrm{eff}}$ exhibits a single peak. Boundary conditions require purely ingoing waves at the event horizon ($r_* \to -\infty$) and outgoing radiation at asymptotic infinity ($r_* \to +\infty$), producing the quasi-normal ringing characterized by exponential decay, as demonstrated in Refs. \cite{Zhang:2024nny,Tang:2025mkk,Tang:2024txx}. 

\begin{figure}[ht!]
	\centering{
		\includegraphics[width=7.3cm]{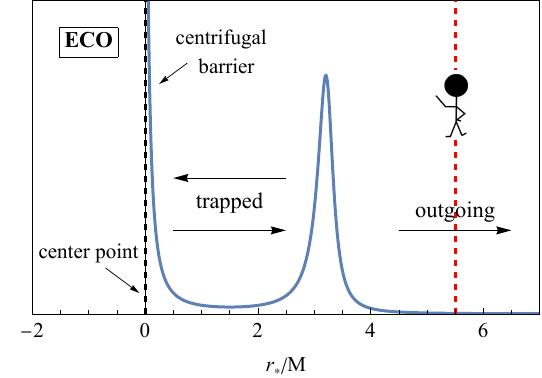}\hspace{4mm}
		\includegraphics[width=8cm]{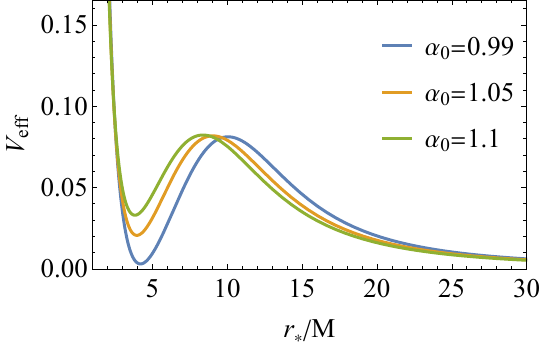}
		\caption{Effective potential profiles. Left panel: Schematic representation of the effective potential for a generic ECO. Right panel: $\alpha_0$-dependent effective potential for PS-ECOs, showing characteristic double-barrier structure.}\label{fig2}
	}
\end{figure}

In contrast to BHs, horizonless compact objects develop a distinctive double-peaked potential barrier structure. This configuration causes incident GWs from infinity to experience partial reflection at the inner centrifugal barrier, producing characteristic echo signals superimposed on the initial burst in the time-domain waveform. The schematic potential profile is illustrated in the left panel of Fig.~\ref{fig2}.

The right panel of Fig. \ref{fig2} shows the behavior of the effective potential for PS-ECO with varying $\alpha_0$. We find that, mirroring the schematic in the left panel, the effective potential exhibits an additional potential barrier near the center, which is a typical signature of GW echoes. When increasing $\alpha_0$, the inter-barrier separation reduces, resulting in a narrower potential well width. This means that the scattering time between the two potential barriers becomes shorter. The complete time-domain analysis follows in the next section.

\section{Numerical Results: echoes and quasinormal modes}\label{sec3}

In this section, we perform a detailed analysis of the characteristic signatures of GW echoes and QNMs of the PS-ECO. To characterize the echo waveforms, we employ time-domain integration techniques as detailed in the Appendix~\ref{appendix-1}. In order to obtain more significant GW signals, we set the angular mode number to $l=1$ and select representative quantum correction parameters $\alpha_0 = \{0.99, 1.05\}$.

\begin{figure}[ht!]
\centering{
\includegraphics[width=7.8cm]{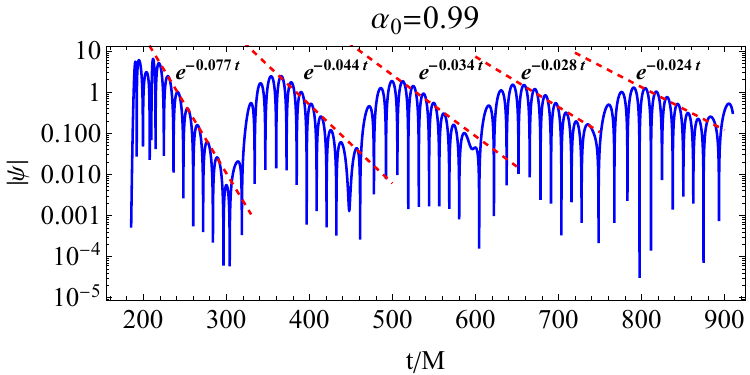}\hspace{0.05mm}
\includegraphics[width=7.8cm]{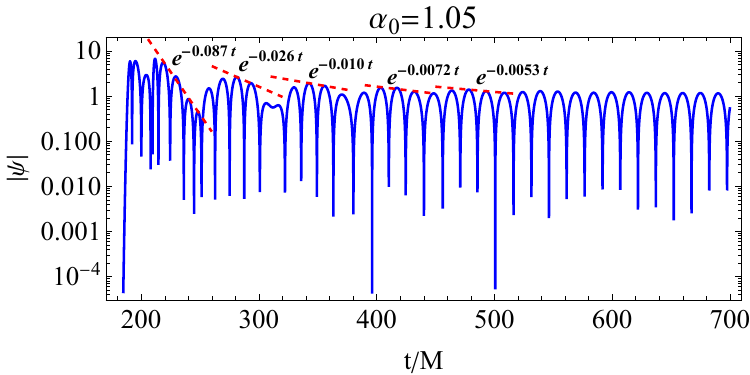}\hspace{0.05mm}
\caption{Semi-log plots of the time evolution of perturbations for $\alpha_0=0.99$ and $\alpha_0=1.05$. }
\label{fig3}}
\end{figure}

As shown in Fig. \ref{fig3}, the time-domain profile of the perturbation for the PS-ECO exhibits significant GW echoes. In the initial ringdown phase, standard damped oscillations are observed, followed by a series of distinct echo signals emerging at intermediate times. 
At late times, the echo signal vanishes, and the evolution is dominated by long-lived modes.
There are several fundamental features that merit highlighting. First, we observed that the echo scattering time $\Delta t$ decreases significantly with increasing $\alpha_0$, consistent with the behavior of the effective potential $V_{eff}$. 
A second key feature is the decay rate of the GW echoes. By performing numerical fits to the time-domain profiles, we determined the decay rate for each individual echo, as represented by the red dashed line in Fig.
\ref{fig3}. The results reveal that later echoes show progressively smaller decay rates compared to preceding ones. For larger $\alpha_0$, the echo signal becomes dominated by long-lived modes more rapidly. Specifically, we have listed the decay exponents for the echoes in Table \ref{tab_2}. 

\begin{table}[ht!]
	\centering
    \begin{tabular}{|c|c|c|c|c|c|}
    \hline  $\alpha_0$  &  \text{initial ringdown} & \text{1st echo} & \text{2ed echo} & \text{3rd echo} & \text{4th echo} \\
    \hline  0.99 &  $0.077$ & $0.044$ & $0.034$ & $0.028$ & $0.024$ \\
    \hline  1.05 &  $0.087$ & $0.026$ & $0.010$ & $0.007$ & $0.005$ \\
    \hline
    \end{tabular}
    \\
\caption{The exponents of the decay for each echoes with $\alpha_0=0.99$ and $1.05$.}  \label{tab_2}
\end{table}

Accurately determining the scattering times from time-domain profiles remains challenging due to the presence of multiple pulse signals. However, the geometric method developed in Ref. \cite{Ikeda:2021uvc} provides a reliable approximate estimate while effectively elucidating the influence of $\alpha_0$ on the scattering dynamics.
From a geometric perspective, the echoes are produced by the wave reflection and transmission between the double potential barriers. The corresponding scattering time can be derived using the geodesic approximation as \cite{Ikeda:2021uvc}
\begin{eqnarray}\label{times}
\Delta t=2 \int_{r_1}^{r_2}\frac{dr}{v_r(r,\alpha_0)}\,, \ \ \ \ v_r(r,\alpha_0)=\frac{dr}{dt}\,,
\end{eqnarray}
where $r_1$ and $r_2$ represent the positions of the two peaks, and $v_r$ denotes the 4-velocity of the photon. We present the scattering time for the representative parameter $\alpha_0$ in Table \ref{tab_3}. It is found that the scattering time decreases as $\alpha_0$ increases, consistent with the behaviors of the time-domain profile shown in Fig. \ref{fig3}. 

\begin{table}[ht!]
	\centering
    \begin{tabular}{ccccccc}
    \hline  $\alpha_0$  &  $\Delta t$ & \text{initial ringdown} & \text{1st echo} & \text{2ed echo} & \text{3rd echo} & \text{4th echo} \\
    \hline  0.99 & 135 & $0.267-0.0779i$ & $0.3006-0.0451i$ & $0.2778-0.0328i$ & $0.2684-0.0271i$ & $0.2594-0.0196i$ \\
    \hline  1.05 & 50 &  $0.2767-0.0873i$ & $0.2488-0.0232i$ & $0.2245-0.0117i$ & $0.2165-0.0070i$ & $0.2089-0.0053i$ \\
    \hline
    \end{tabular}
    \\
\caption{For representative values of $\alpha_0$, scattering times $\Delta t$ derived from geometric approximation (column 2) and QNM frequencies $\omega$ extracted via Prony method (columns 3--7).}  \label{tab_3}
\end{table}

Meanwhile, we perform the Prony method to extract the characteristic frequency of echoes from the temporal waveform data. It should be emphasized that it is difficult for the Prony method to identify the QNMs in this case, due to the superposition of multiple modes. However, this method can help us to understand the characteristics of a single echo, such as the decay rate for a single echo. To this end, the waveform is modeled as a superposition of $p$ exponentially damped components:  
\begin{eqnarray}\label{prony}
\psi(r,t)\simeq \sum_{j=1}^{p}C_j e^{-i \omega_j t}\,,
\end{eqnarray}
where $\omega_j$ denotes the complex characteristic frequency. Using the Prony method, we determine the characteristic frequencies for both the initial outburst and each echo, as cataloged in Table \ref{tab_3}. The initial ringdown is characterized by the most rapid exponential decay rate. Subsequently, the evolution transitions to periodic echo oscillations. Within this regime, the first echo demonstrates the largest oscillation frequency and damping rate, while later echoes exhibit progressively slower damping rates and lower frequencies. This result exhibits a good agreement with the numerical fitting in time-domain profiles (see Fig. \ref{fig3} and Table \ref{tab_2}).

Determining the QNMs remains essential, as they provide a deeper insight into the fundamental characteristics of spacetime geometry. To calculate the QNMs for the PS-ECO, we adopt the direct integration method and provide a brief introduction in the Appendix \ref{appendix-2}. Figs.
\ref{fig4} and \ref{fig4b} show the dependence of the QNM frequencies on the parameter $\alpha_0$. Crucially, as $\alpha_0$ approaches the critical value $0.981$, both components of the fundamental QNM frequency tend toward zero. The imaginary part rapidly tends toward zero (Fig. \ref{fig4} and the right panel of Fig. \ref{fig4b}), indicating the dominance of long-lived modes at late times. The real part approaches zero at a slower rate (Fig. \ref{fig4} and left panel of Fig. \ref{fig4b}), corresponding to a weakly oscillatory behavior. The characteristics of the long-lived modes with weak oscillations provide a definitive spectral signature distinguishing PS-ECOs from BHs. When $\alpha_0$ increases, both the real and imaginary parts of the fundamental mode grow as shown in Fig. \ref{fig4b}. Consequently, the late-time evolution transitions from long-lived modes to oscillatory decay behavior. 
For higher overtone modes, the QNMs do not exhibit the overtone outburst, which is completely different from the results observed for BHs in Ref. \cite{Zhang:2022hxl,Tang:2024txx,Gong:2023ghh,Fu:2023drp,Konoplya:2022hll,Konoplya:2022pbc,Konoplya:2022iyn,Konoplya:2023ppx,Zhu:2024wic,Tang:2025mkk}. Notably, the QNM spectrum does not exhibit a smooth transition from BH to PS-ECO, but rather manifests a distinctly different set of modes, including long-lived modes. This indicates that the frequency exhibits non-perturbative behavior in $\alpha_0$, which stems completely from the modified boundary conditions at the horizon.

\begin{figure}[ht]
	\centering{
		\includegraphics[width=7.8cm]{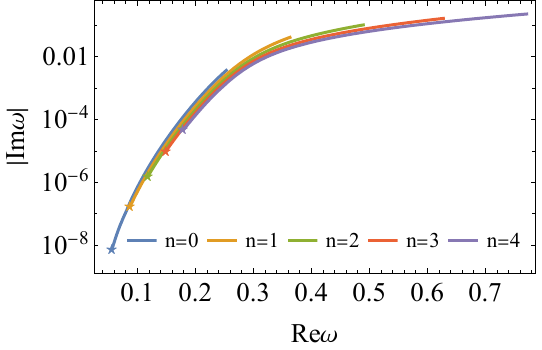}\hspace{1mm}
		\caption{The QNM frequencies spectrum  with fundamental modes and overtones for $\alpha_0$ from $0.985$ to $1.1$. The star markers indicate QNM values corresponding to $\alpha_0=0.985$. }
		\label{fig4}
	}
\end{figure}

\begin{figure}[ht]
	\centering{
		\includegraphics[width=7.5cm]{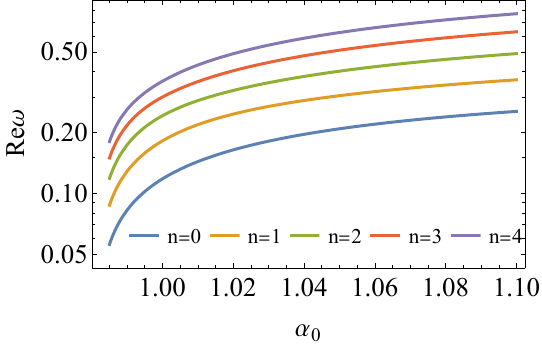}\hspace{4mm}
		\includegraphics[width=7.8cm]{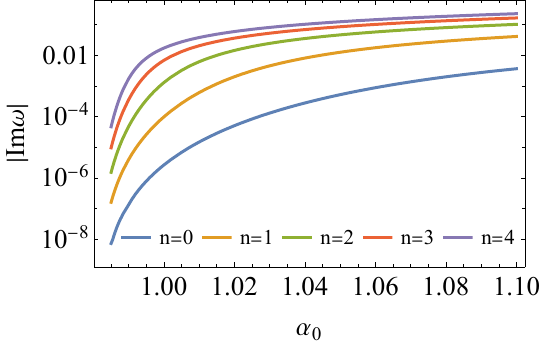}
		\caption{The QNMs as a function of the $\alpha_0$. The left panel represents the real parts of the QNMs, while the right panel shows the absolute values of their imaginary parts.}
		\label{fig4b}
	}
\end{figure}

\begin{figure}[ht]
	\centering{
		\includegraphics[width=8.6cm]{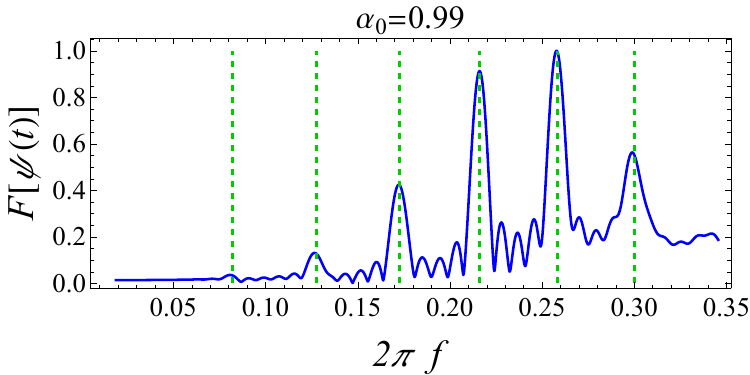}\hspace{8mm}
		\includegraphics[width=5cm]{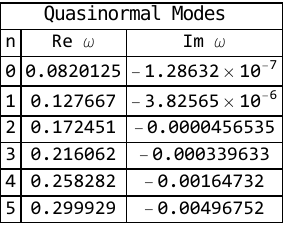}
		\caption{The Fourier spectrum with the $\alpha_0=0.99$ in the frequency domain (left panel) and its corresponding QNMs (right panel). The green line represents the real part of the QNMs in the right panel. }
		\label{fig5}
	}
\end{figure}

To better understand the results of the time evolution and QNMs, we employ the discrete Fourier transform to convert time-domain profiles into frequency-domain spectra \cite{Tan:2023cra,Zhu:2023tzx}. The discrete Fourier transform is 
\begin{eqnarray}\label{DFT}
F \left[\psi(t)\right]=\big| \Sigma_{p}\psi(t_p,z)\exp(-2\pi ift_p)\big|\,,
\end{eqnarray}
where $t_p$ is the discrete time. Applying Eq.~\eqref{DFT} to the waveforms corresponding to $\alpha_0=0.99$ in the left panel of Fig.~\ref{fig3} yields the Fourier spectrum shown in the left panel of Fig.~\ref{fig5}. The spectrum exhibits several resonance peaks, with sequential peaks (left to right) corresponding to the real parts of the fundamental QNM and its higher overtones. This one-to-one mapping demonstrates excellent consistency between the time-domain evolution and the QNM spectrum.

\section{Conclusions and discussions}\label{sec4}

In this paper, we extend the RBH model proposed in \cite{Ling:2021olm} to describe horizonless ultracompact stars, which are classified based on photon sphere topology as either PS-ECOs or PL-ECOs. We focus on the GW echoes and QNMs of the PS-ECO—a compact object featuring a Minkowski-core enclosed by a photon sphere.

A key finding is that the effective potential for electromagnetic perturbations in the PS-ECO model features a distinct inner potential barrier near the center, markedly distinct from BH potentials. This structure is responsible for the GW echo phenomenon. Notably, quantum gravity effects reduce the inter-barrier separation, resulting in a narrower potential well width and a corresponding decrease in the scattering time between barriers.

A detailed time-domain analysis reveals that the dynamical evolution of perturbed PS-ECOs can be divided into three consecutive phases: the initial ringdown, the echo phase, and the final ringdown. Their characteristics are summarized as follows:
\begin{itemize}
    \item The initial-ringdown phase
    \begin{itemize}
    \item The initial-ringdown phase exhibits the standard damped oscillations characterized by the most rapid exponential decay rate.
\end{itemize}
\item The echo phase
    \begin{itemize}
    \item This phase is characterized by periodic echo signals. The first echo has the highest frequency and damping rate, with subsequent echoes showing progressively reduced frequencies and slower decay.
    \item Both the analysis in the time-domain and the approximate estimate using the geometric method confirm that quantum gravity effects significantly reduce the echo time delay, consistent with the evolution of the effective potential.
\end{itemize}
\item The final-ringdown phase
    \begin{itemize}
    \item Echoes eventually vanish, giving way to long-lived modes with weak oscillations. These modes provide a clear spectroscopic signature distinguishing PS-ECOs from BHs.
    \item Higher overtones do not exhibit the typical ``overtone outburst" behavior observed in BH ringdown signals.
    \item Quantum gravity effects accelerate the dissipation of echoes, prompting an earlier transition into this phase dominated by the long-lived mode.
\end{itemize}
\end{itemize}

Another fundamental result is that the QNM spectrum of this PS-ECO does not connect smoothly to that of the BH. Instead, the PS-ECO supports a completely different set of modes—including long-lived ones—indicating non-perturbative behavior in the quantum correction parameter $\alpha_0$.   

This work provides clear and discriminative features in the echo properties and late-time ringdown that may help identify ECOs with future GW detectors. The study highlights how quantum corrections alter the effective potential and dynamical evolution of perturbations, particularly in accelerating echo damping. The absence of overtone outbursts and the presence of long-lived modes underline fundamental differences in the dynamics of horizonless ultracompact objects compared to BHs.

The current model is restricted to electromagnetic perturbations; including gravitational perturbations will be essential for realistic GW comparisons. Once the gravitational perturbations are included, future work could also connect these results with actual observational constraints or detection strategies for ECOs with instruments like LIGO-Virgo-KAGRA or future space-based detectors. In addition, the analysis assumes a specific quantum-gravity-inspired model. Testing against other models of quantum corrections or horizonless objects would strengthen the conclusions.

In summary, this study offers a coherent picture of the ringdown and echo features of a quantum-corrected horizonless compact object. It notifies a clear imprint of quantum effects in the waveform and provides a theoretical foundation for future searches of exotic compact objects beyond BHs.

\acknowledgments

We are sincerely grateful to Li Shulan and Yuan Tianbao for their valuable discussions.
This work is supported by National Key R$\&$D Program of China (Grant No. 2020YFC2201400), the Natural Science Foundation of China (Grants Nos. 12505085, 12447151, 12375055, 12347159, 12275079, 12405055, 12347111, 12035005, 12447156 and 12505078). It was also supported by the China Postdoctoral Science Foundation (Grant Nos. 2025T180931, 2025M773339 and 2023M741148), the Postdoctoral Fellowship Program of CPSF (Grant No. GZC20240458), the Postgraduate Scientific Research Innovation Project of Hunan Province under Grant No. CX20220509, and the Jiangsu Funding Program for Excellent Postdoctoral Talent (Grant No. 2025ZB705).
	
\appendix
	
\section{The numerical methods}\label{method}
This appendix provides a brief overview of the numerical methods employed in this work: the time-domain integration and the direct integration method. 

\subsection{Time-domain integration}\label{appendix-1}
To explore the dynamics evolution of the electromagnetic perturbation, we implement the finite difference method on a uniform grid with discrete intervals $\{\Delta t, \ \Delta r_*\}$ \cite{Abdalla:2010nq,Zhu:2014sya,Lin:2022rtx}. The function $\psi^{em}(t,r_*)$ and the effective potential $V_{eff}(r(r_*))$ are discretized as
\begin{equation}
\psi^{em}(t,r_*)=\psi^{em}(i \Delta t,j\Delta r_*)=\psi_{i,j} \,, \ V(r(r_\ast))=V(j\Delta r_\ast)=V_j\,,
\end{equation}
where the indices $i$ and $j$ denote the number of time and space grid points, respectively. Applying this discretization scheme to Eq. \eqref{eq_Sch} yields the finite difference equation:
\begin{equation}
	\begin{split}
		-\frac{(\psi_{i+1,j}-2\psi_{i,j}+\psi_{i-1,j})}{\Delta t^2}+\frac{(\psi_{i,j+1}-2\psi_{i,j}+\psi_{i,j-1})}{\Delta r^2_\ast}-V_{j}\psi_{i,j}=0.
	\end{split}
\end{equation}
Subsequently, the finite difference iteration equation can be expressed as  
\begin{equation}
	\psi_{i+1,j}=-\psi_{i-1,j}+\frac{\Delta t^2}{\Delta r^2_\ast}(\psi_{i,j+1}+\psi_{i,j-1})+(2-2\frac{\Delta t^2}{\Delta r^2_{\ast}}-\Delta t^2V_{j})\psi_{i,j}.
\end{equation}
The stability of the evolution requires $\Delta t/\Delta r_*<1$.  It is crucial to emphasize that reliable numerical results depend not only on the ratio $\Delta t/\Delta r_\ast$ but also on the individual values of $\Delta t$ and $\Delta r_\ast$. Furthermore, taking into account the quasinormal ringing is independent of the initial perturbation, we consider an initial Gaussian distribution $\psi(t=0,r_\ast)=\mathrm{exp}[-\frac{(r_\ast-\bar{a})^2}{2b^2}]$ with $\psi(t<0,r_\ast)=0$. For the horizonless compact object, the perturbation must vanish at the center, corresponding to a Dirichlet condition at $r_*=0$ as 
\begin{equation}\label{BC_0}
\psi(t,r_*=0)=0\,.
\end{equation}

\subsection{Direct Integration Method}\label{appendix-2}

For a given system, various approaches can be utilized to obatin QNM frequencies, including but not limited to the WKB approximation \cite{1985ApJ-291L33S,Iyer:1986np,Guinn:1989bn,Konoplya:2004ip,Konoplya:2003ii,Matyjasek:2017psv}, the spectral method \cite{Boyd:Chebyshev,Jansen:2017oag,Fu:2022cul,Fu:2023drp,Gong:2023ghh,Zhang:2024nny}, and the continued fraction method (CFM) \cite{Leaver:1985ax,Pani:2013pma}, with the selection determined by both the specific physical problem at hand and the respective strengths and shortcomings of the methods themselves. In this work, we used the direct integration method \cite{Ferrari:2007rc,Pani:2012bp,Pierini:2023btw}, which has the advantage of being adaptable to the boundary conditions of different systems while also extendable to the coupled differential equations. We begin by considering a single second-order wave equation:
\begin{equation}
\psi''(r)+A_1(r)\psi'(r)+B_1(r)\psi(r)=0\,,
\end{equation}
with
\begin{equation}
A_1(r)=f'(r)/f(r)\,, \ \ B_1(r)=(\omega^2-V_{eff}(r))/f(r)^2\,.
\end{equation}

For the QNM analysis, the boundary condition at infinity is a purely outgoing wave,
\begin{equation}
\psi_{out}(r\sim \infty)=\sum_{i=0}^{N_\infty} e^{-i\omega r_*} \psi_i^{\infty}(\omega)/r^{i}\,,
\end{equation}
where the truncation order $N_{\infty}$ must be chosen sufficiently large to counteract the numerical inaccuracies introduced by the finite cutoff at infinity. However, near the center, we impose the Dirichlet boundary condition as 
\begin{equation}
\psi_{in}(r\sim 0)=0\,.
\end{equation}
which is consistent with the boundary condition \eqref{BC_0} in the time-domain evolution.

\bibliographystyle{style1}
\bibliography{Ref}
\end{document}